\def\epsfpreprint{Y}   

                       \if \epsfpreprint Y
\documentstyle[12pt,epsf]{article}
\fi \if
                       \epsfpreprint N
\documentstyle[12pt]{article} \fi

\def\figsizeA{4.8}

\def\figure#1#2#3{\if \epsfpreprint Y \epsfxsize=#3truein
\centerline{\epsffile{fig_#1.eps}}
\centerline{\vbox{{\bf \noindent Figure #1.} #2}}
\bigskip \fi}

\def\Psibar{\overline{\Psi}}
\def\slash{\!\!\!/}
\def\bfm#1{\mbox{\boldmath $#1$}}
\def\MeV{{\rm\  MeV}}
\def\GeV{{\rm\  GeV}}

\def\spose#1{\hbox to 0pt{#1\hss}}
\def\ltapprox{\mathrel{\spose{\lower 3pt\hbox{$\mathchar"218$}}
 \raise 2.0pt\hbox{$\mathchar"13C$}}}
\def\gtapprox{\mathrel{\spose{\lower 3pt\hbox{$\mathchar"218$}}
 \raise 2.0pt\hbox{$\mathchar"13E$}}}
\def\inapprox{\mathrel{\spose{\lower 3pt\hbox{$\mathchar"218$}}
 \raise 2.0pt\hbox{$\mathchar"232$}}}

\def\one{$r=0$, ${M_\pi \over F_\pi} = {140 \MeV \over 93 \MeV}$ and
the number of colors is set equal to three. From top to bottom the
lines correspond to the $\sigma$ mass, quark mass and $\sigma$ width
calculated at large $N$ and to leading order in $m_\sigma$. The
vertical line denotes the point where ${M_q \over F_\pi} = { 310 \MeV
\over 93 \MeV }$.}
\def\two{The $\beta_1$, $\kappa = 1 / (8 +2 m_0)$ plane for $r=1$,
$N=2$. The solid lines are constant $m_\pi$ lines.  From top to
bottom they correspond to $m_\pi=0, 0.1, 0.2, 0.3$. The dotted lines
are constant $m_q$ lines. From right to left they correspond to
$m_q=-0.2,-0.1,0.0$, $0.1$, $0.2$, $0.3$. The $m_q=0, m_\pi=0$ point
is located at $\beta_{1_{chiral}} =0.3416$, $\kappa_{chiral}=0.3994$.}
\def\three{$r=0$, $m_0=0$, $N=2$.  The diamonds are the values of
$m_q$ and the crosses are the values of $m^{\prime}_q$ from the
numerical simulation for $lx=8$, $lt=16$.  The solid lines are the
large $N$ numbers with the zero mode included. From right to left they
correspond to $(L_x=8,L_t=16)$, $(L_x=16,L_t=16)$, $(L_x=32,L_t=32)$,
$(L_x=64,L_t=64)$. The dotted lines are the large $N$ numbers with the
zero mode excluded. From right to left they correspond to
$(L_x=16,L_t=16)$, $(L_x=8,L_t=16)$.  The solid vertical line denotes
the infinite volume $\beta_{1_c}$ from large $N$. For
$\beta_1\le\beta_{1_c}$ the model is in the broken phase.}
\def\four{ {\bf a)} The $\sigma$ propagator in momentum space for $10$
small momenta with $r=0$, $m_0=0$, $\langle \sigma \rangle = 0.4840$,
$N=2$, $L_x=16$, $L_t=16$. The crosses are the values from the
numerical simulation.  The diamonds are the large $N$ results. {\bf
b)} Same as in {\bf a} but for $\langle \sigma \rangle = 0.35$.  {\bf
c)} Same as in {\bf a} but for the pion propagator. {\bf d)} Same as
in {\bf b} but for the pion propagator.}
\def\five{$r=1$, $N=2$, $L_x=8$, $L_t=16$. The crosses are the MC
data. The solid line is the large $N$ prediction on the same size
lattice.}
\begin{document}

\title{\bf The Nambu--Jona-Lasinio Model of QCD on the Lattice}
\vskip 1. truein
\author{Khalil M. Bitar and Pavlos M. Vranas \\
Supercomputer Computations Research Institute \\
The Florida State University \\
Tallahassee, FL 32306 \\ \\ \\
FSU--SCRI--93--127}
\maketitle

\vskip 1.5 truein

\begin{abstract}

In an effort to investigate some of the low energy properties of QCD,
in particular those related to chiral symmetry breaking, as well as to
obtain insights on the behavior of an interacting theory of fermions
on the lattice, the two flavor Nambu--Jona-Lasinio model with $SU(2)
\times SU(2)$ chiral symmetry is studied on the four--dimensional
hypercubic lattice using large $N$ techniques and numerical
simulations. Naive and Wilson fermions are considered and transparent
results are obtained regarding the following: the scalar and
pseudoscalar spectrum, the approach to the continuum and chiral
limits, the size of the $1/N$ corrections, and the effects of the zero
momentum fermionic modes on finite lattices. Also, some interesting
observations are made by viewing the model as an embedding theory of
the Higgs sector.

\end{abstract}

\newpage

When the high frequency modes of the gauge and fermionic fields of QCD
are integrated down to the energy scale $E$ corresponding the
correlation length of the gauge field ($E \approx$ glueball mass
$\approx 1550 \MeV$ \cite{UKQCD}) the resulting effective theory will
essentially be a theory of fermions with contact interactions and
cutoff $\Lambda \ltapprox 1550 \MeV$. The resulting effective
Lagrangian will maintain the original chiral symmetry but will be more
complicated. At energies much smaller than $\Lambda$ it should be
enough to keep in the Lagrangian the least irrelevant operator, namely
the four--Fermi dimension six operator. This is one way
\cite{Dhar-Shankar-Wadia} to motivate the Nambu--Jona-Lasinio (NJL)
model.

Unfortunately, by only keeping the four--Fermi operator, valuable
information was lost and the NJL model does not confine the quarks.
Furthermore, if, for example, we want to study the $\sigma$ particle,
which on phenomenological grounds is believed to have mass $\approx
750 \MeV$, then the separation of scales is probably not large enough
to justify the neglect of operators with dimension higher than six.
Nevertheless, the NJL model possesses the same chiral symmetry as QCD
and it can realize this symmetry in the Goldstone mode. It is this
feature that is most crucial in the understanding of the lightest
hadrons. Furthermore, our interest in the model is not so much aimed
at its quantitative predictability but rather on the qualitative
insights that can provide as a low energy theory of QCD, as an
embedding theory of the Scalar Sector of the Minimal Standard Model
and as a four--dimensional interacting theory of fermions on the
lattice.

The NJL model has been studied extensively for various cases with
continuum type regularizations. For a comprehensive review the reader
is referred to \cite{Klevansky} and references therein. The model has
also been studied on the lattice \cite{Hasenfratz} in connection with
the possible equivalence of the top quark condensate with the Higgs
field \cite{Bardeen}. In that work, however, the separation of scales
is very large (the cutoff is of the order $10^{14} \GeV$), and it is
therefore quite a different problem than the one considered here.

In this paper we consider the two flavor (up and down) NJL model with
$SU(2) \times SU(2)$ chiral symmetry and SU(N) color symmetry, with
scalar and pseudoscalar couplings \cite{Ebert-Volkov} on the
four--dimensional hypercubic lattice. We consider both naive and
Wilson fermions and we study the model using a large $N$ expansion as
well as a Hybrid Monte Carlo (HMC) \cite{DKPR} numerical simulation.
In this paper we will only present the main results. A detailed
version of this work will appear elsewhere \cite{Bitar-Vranas}.

The Lagrangian density in Minkowski space and in continuum notation
is:
\begin{equation}
{\cal L} = \Psibar(i \partial\slash  - m_0)\Psi +
{G_1 \over 2}\left[(\Psibar\Psi)^2 +
(\Psibar i \gamma_5 \bfm{\tau} \Psi)^2 \right].
\label{Langrangian1}
\end{equation}
In the above expression all indices have been suppressed. The fermionic
field $\Psi$ is a flavor $SU(2)$ doublet and a color $SU(N)$
$N$-column vector. The Lagrangian is diagonal in color, in contrast
with the full QCD Lagrangian which is diagonal in flavor. $\bfm{\tau}
=\{\tau_1,\tau_2,\tau_3\}$ are the three isospin Pauli matrices,
$\partial\slash = \gamma^\mu \partial_\mu$, and $m_0$ is the bare
quark mass (if $m_0 \neq 0$ the chiral symmetry is explicitly broken).
To obtain a Lagrangian that is quadratic in the fermionic fields we
introduce the scalar auxiliary field $\sigma$ and the three
pseudoscalar auxiliary fields $\bfm{\pi}=\{\pi_1, \pi_2, \pi_3\}$
\cite{Ebert-Volkov}.

Because of the chiral couplings the fermionic determinant has a phase.
This phase is related to the chiral anomaly and the Wess--Zumino term
\cite{Ebert-Reinhard} and we will not consider it in this work.  Going
to Euclidean space and appropriately discretizing the above Lagrangian
we obtain the model on the hypercubic lattice. On the lattice, as it
is well known, we have species doubling. The doubling in the NJL model
will be interpreted as a doubling of the color degrees of freedom. To
treat this problem we add to the Lagrangian density an irrelevant
operator (Wilson term) of the form ${a r \over 2} \Psibar \partial^2
\Psi$, $a$ being the lattice spacing and $r$ a constant. We consider
the $r=0$ case where no effort is made to remove the doublers (naive
fermions) and also the $r \neq 0$ case where the doubler masses are
raised to the cutoff and the chiral symmetry is explicitly broken
(Wilson fermions). With these considerations and after appropriate
scaling of the fields and couplings, so that only dimensionless
quantities appear, we obtain:
\begin{eqnarray}
Z &=& \int[d\Psi
d\Psibar d\sigma d\bfm{\pi}] e^{-S} \nonumber \\
S &=& \sum_{x,y}
\left\{ \sum_{i=1}^{N/2} \left\{\Psibar^i_x M_{xy} \Psi^i_y +
\Psibar^{i+N/2}_x M^\dagger_{xy} \Psi^{i+N/2}_y \right\} + n_f \beta_1
(\sigma^2_x + \bfm{\pi}^2_x) \delta_{xy} \right\}
\nonumber \\
M_{xy} &=& {1\over 2} \sum_{\mu} \left[
(\gamma_\mu-r)\delta_{x+\mu,y} -
(\gamma_\mu+r)\delta_{x-\mu,y} \right] + (4 r + m_0 + \sigma_x + i
\gamma_5 \bfm{\pi_x\cdot\tau} )\delta_{xy}
\label{PartFun2}
\end{eqnarray}
with $\gamma_\mu$ hermitian, $n_f=2$ the number of flavors and
$\beta_1 = {1 \over 2 n_f G}$. For $r=0$ and $m_0=0$ the symmetry is
not explicitly broken. In that case at $\beta_1 = \beta_{1_c}$ the
model undergoes a second order phase transition from a symmetric phase
with $\langle \sigma \rangle =0 $, massive $\bfm{\pi}$ and $\sigma$
fields and massless quarks, to a phase with spontaneously broken
symmetry with $\langle \sigma \rangle \ne 0$, massless pion fields
(Goldstone particles) and massive $\sigma$ and quark fields with
dynamically generated quark mass equal to $\langle \sigma \rangle$. In
this work we will always stay in the broken phase.

We study the above action using large $N$ techniques and obtain
analytical results both on finite and infinite volumes. The infinite
volume results are obtained sufficiently close to the continuum limit
using asymptotic expansions. We also study this action for $N=2$ using
an HMC numerical simulation with Conjugate Gradient and leap--frog
algorithms. The seven important results that stem from our analysis
are presented below.

\medskip
\noindent {\bf I}
\smallskip

For naive fermions, at large $N$ and with pion mass $M_\pi=140 \MeV$,
we calculate the $\sigma$ mass ($M_\sigma$), the $\sigma$ width
($\Gamma_\sigma$) and the quark mass ($M_q$) in physical units as
functions of the $\sigma$ mass $m_\sigma$ in lattice units. The
results are given in Figure 1 in units of the pion decay constant
$F_\pi = 93 \MeV$. By setting $M_q = 310 \MeV$ we find $M_\sigma = 726
\MeV$, $\Gamma_\sigma=135 \MeV$, and $\Lambda = \pi / a = 1150 \MeV$
(dotted vertical line). $M_\sigma$ is consistent with phenomenological
expectations and $\Lambda$ is consistent with the expectation that the
cutoff should be close and below the mass of the lightest glueball
($1550 \MeV$). The width however is underestimated.  The reason is
traced to the fact that to leading order in large $N$ the width
receives contributions only from the quark bubble and not from the
pion bubble because the pion bubble is of order $1/N$.  Because the
phase space available for the $\sigma$ to decay to two quarks is much
smaller than the phase space to decay to two pions, the pion loop
contribution, although of order $1 / N$, is probably more important
than the quark loop contribution.

\medskip
\noindent {\bf II}
\smallskip

The above result is relevant not only for the low energy QCD but also
for the Higgs sector. It is well known that there is an equivalence
between the $\sigma$-$\pi$ sector of QCD with the scalar sector of the
Minimal Standard Model.  In the former the scale is set by the pion
decay constant ($F_\pi=93 \MeV$) and in the latter by the weak scale
($F_\pi = 246 \GeV$). As mentioned in {\bf I} we find that in
accordance with phenomenological expectations in the $\sigma$-$\pi$
QCD sector $M_\sigma / F_\pi \approx 8$ but in the Higgs sector all
previous analysis predicts a triviality bound of the Higgs mass with
$M_\sigma / F_\pi \ltapprox 2.8$ (see for example \cite{HKNV}). In the
past this has been a reason for concern since it could imply that the
Higgs mass bound may be underestimated. Our analysis suggests that
this apparent discrepancy appears because the Higgs mass bounds are
traditionally obtained for $m_\sigma \ltapprox 0.5$ while the
$M_\sigma / F_\pi \approx 8$ ratio is obtained for $m_\sigma \approx
2$ and it should therefore be accompanied by large deviations from the
low energy behavior. Nevertheless, this is only a suggestion since we
have not calculated deviations from the low energy behavior of a
physical process that would enable us to make exact statements.
However, the value of the width serves as an indication of the size of
such deviations. In a way, departure from low energy behavior will be
signaled by an increasing width of the $\sigma$ to two quark decay. At
$m_\sigma \approx 2$ the width is already fairly large.

\figure{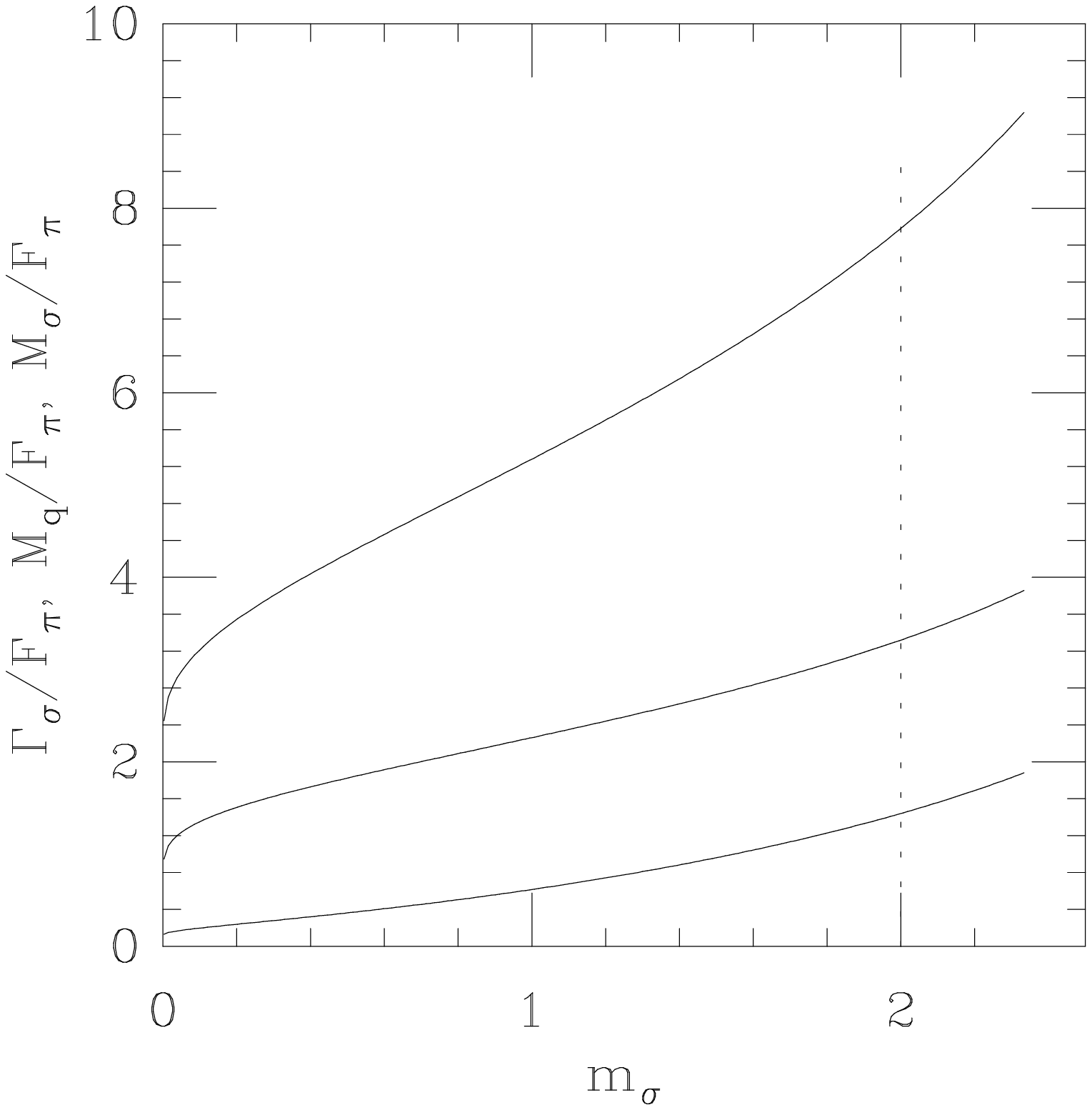}{\one}{\figsizeA}

\medskip
\noindent {\bf III}
\smallskip

If the Higgs sector is the low energy effective field theory of a NJL
model with $N_c=3$, $n_f=2$ and $M_\pi=0$ (the pions should now be
viewed as the would--be Goldstone Bosons), we obtain for the larger
values of $m_\sigma$ a figure that is almost exactly the same as
Figure 1 with ($M_\sigma$), ($\Gamma_\sigma$) and ($M_q$) measured in
units of $F_\pi=246 \GeV$.  Strictly speaking such a figure does not
contain any physical predictions but there is a very interesting
observation that can be made.

If we set the fermion mass to ${M_q \over F_\pi} \approx {310 \over
93}$, as is the case for the low energy sector of QCD, then the Higgs
mass will be $M_\sigma = 1915\GeV$.  This point will correspond to
$m_\sigma = 2$ as denoted by the dotted vertical line in Figure 1. As
discussed in {\bf II}, at $m_\sigma = 2$ one would expect very large
deviations from the low energy behavior of scattering cross sections.
This suggests that as the CM energy is turned up first the deviations
from the low energy behavior will become large, signaling the onset of
new physics, and later the Higgs particle would be observed.

\figure{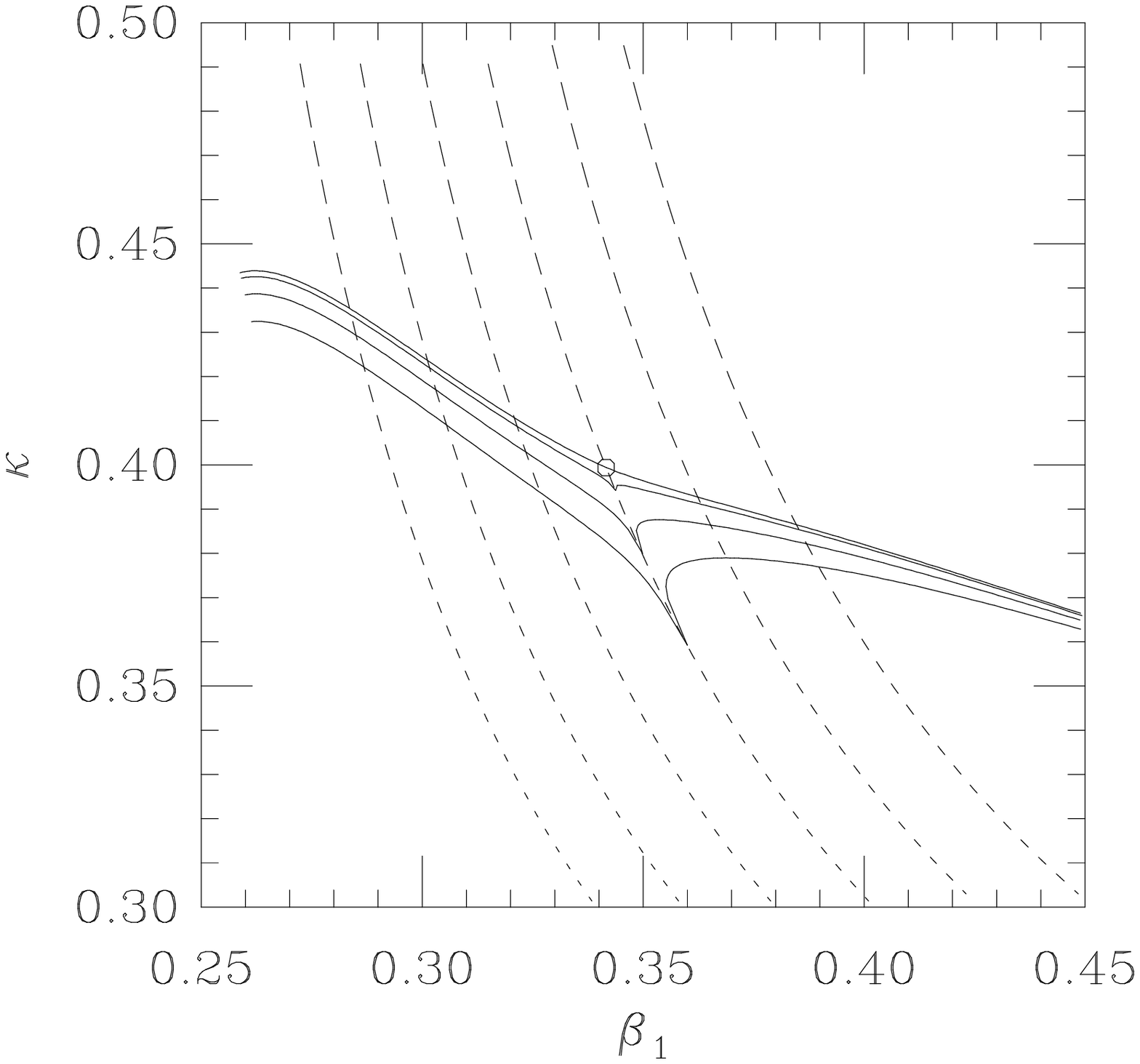}{\two}{\figsizeA}

\medskip
\noindent {\bf IV}
\smallskip

With Wilson fermions we obtain at large $N$ analytical expressions of
the pion mass ($m_\pi$) and quark mass ($m_q$) in lattice units as
functions of the bare parameters of the model. We are then able to
make exact statements regarding the approach to the continuum and
chiral limits. For small $m_q$ and $m_\pi$ the constant $m_q$ and
$m_\pi$ lines including the $m_q=0$ and $m_\pi=0$ lines are presented
in Figure 2. The point $\beta_{1_{chiral}}$, $\kappa_{chiral}$, where
the $m_q=0$ and $m_\pi=0$ lines intersect, is denoted by a circle and
corresponds to the true chiral limit.

{}From this figure we see that if for a fixed $\beta_1$ we were to
change $\kappa={1\over 8r + m_0}$ from smaller to larger values (as is
often done in QCD with dynamical Wilson fermions) then if $\beta_1 <
\beta_{1_{chiral}}$ we would reach the $m_\pi =0$ limit before we
reach the continuum limit. On the other hand if $\beta_1 >
\beta_{1_{chiral}}$ we would reach the continuum limit before we reach
the $m_\pi =0$ limit. As mentioned above there is only one point in
the $\beta_1$, $\kappa$ plane where we can obtain a true chiral limit.
This may provide an insight on how the retrieval of the chiral limit is
achieved in QCD.

\figure{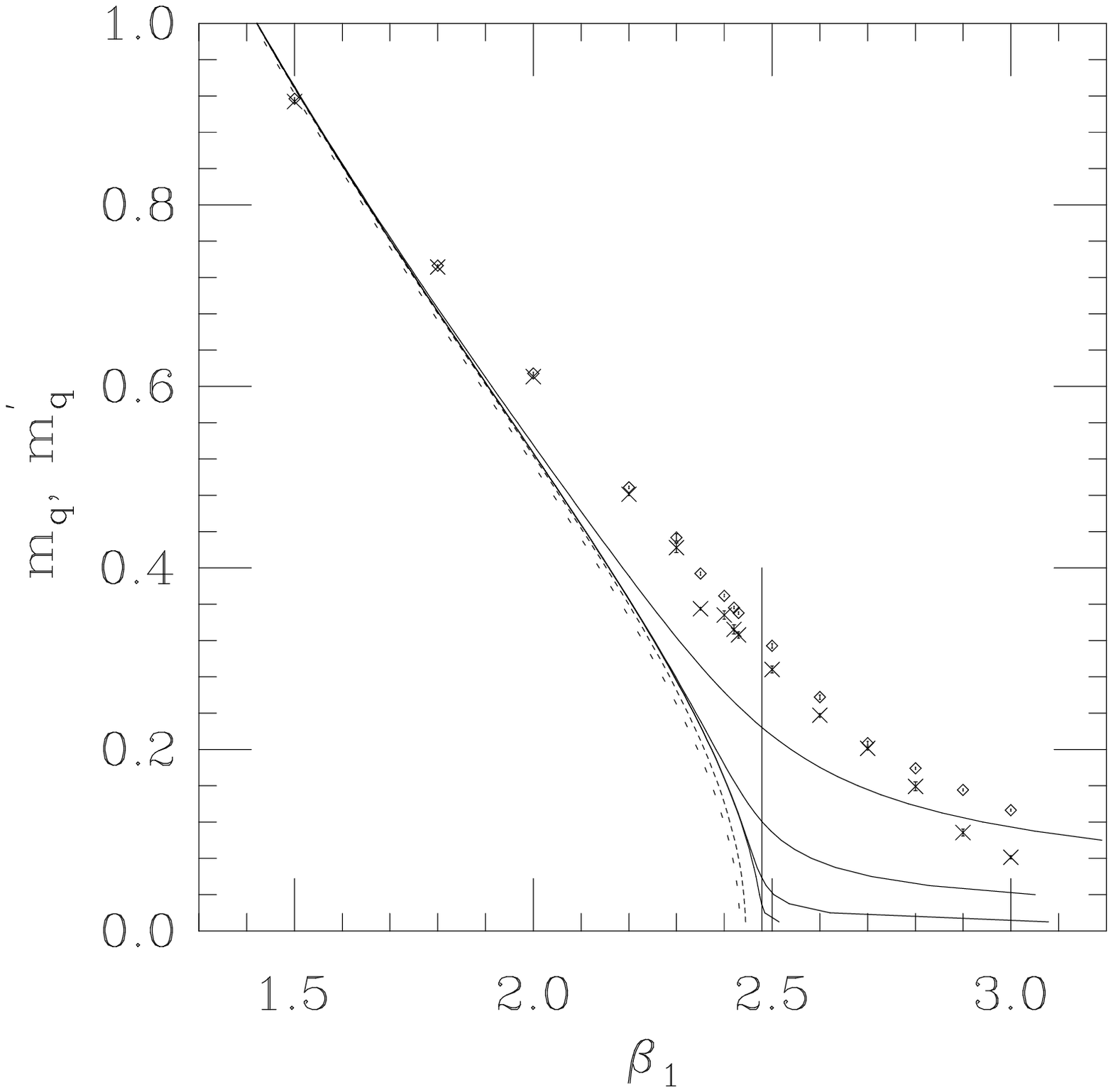}{\three}{\figsizeA}

In Figure 2 the constant $m_\pi$ lines were drawn under the assumption
that $m_\pi^2 \ll 4 m_q^2$ which can not be satisfied on the $m_q =0$
line, except on the one point where it intersects the $m_\pi=0$ line.
The constant $m_\pi$ lines are therefore valid only in the regions
where $m_\pi^2 \ll 4 m_q^2$.

\medskip
\noindent {\bf V}
\smallskip

At large $N$ and for Wilson fermions the $\sigma$ particle has mass
proportional to the cutoff. Our analysis traces this fact to two
related reasons. First, although the Wilson term has raised the masses
of the doublers to the cutoff, it has not decoupled them from the
theory. Through vacuum polarization these contribute to the $\sigma$
self energy and raise its mass. Second, although the Wilson term has
not altered the low frequency behavior of the propagating quark, it
has however altered its high frequency behavior. Again through vacuum
polarization the high frequency modes contribute to the $\sigma$ self
energy and also raise its mass. Such a phenomenon may also be
responsible for the difficulty in observing the $\sigma$ particle in
numerical simulations of QCD with dynamical Wilson fermions.

\figure{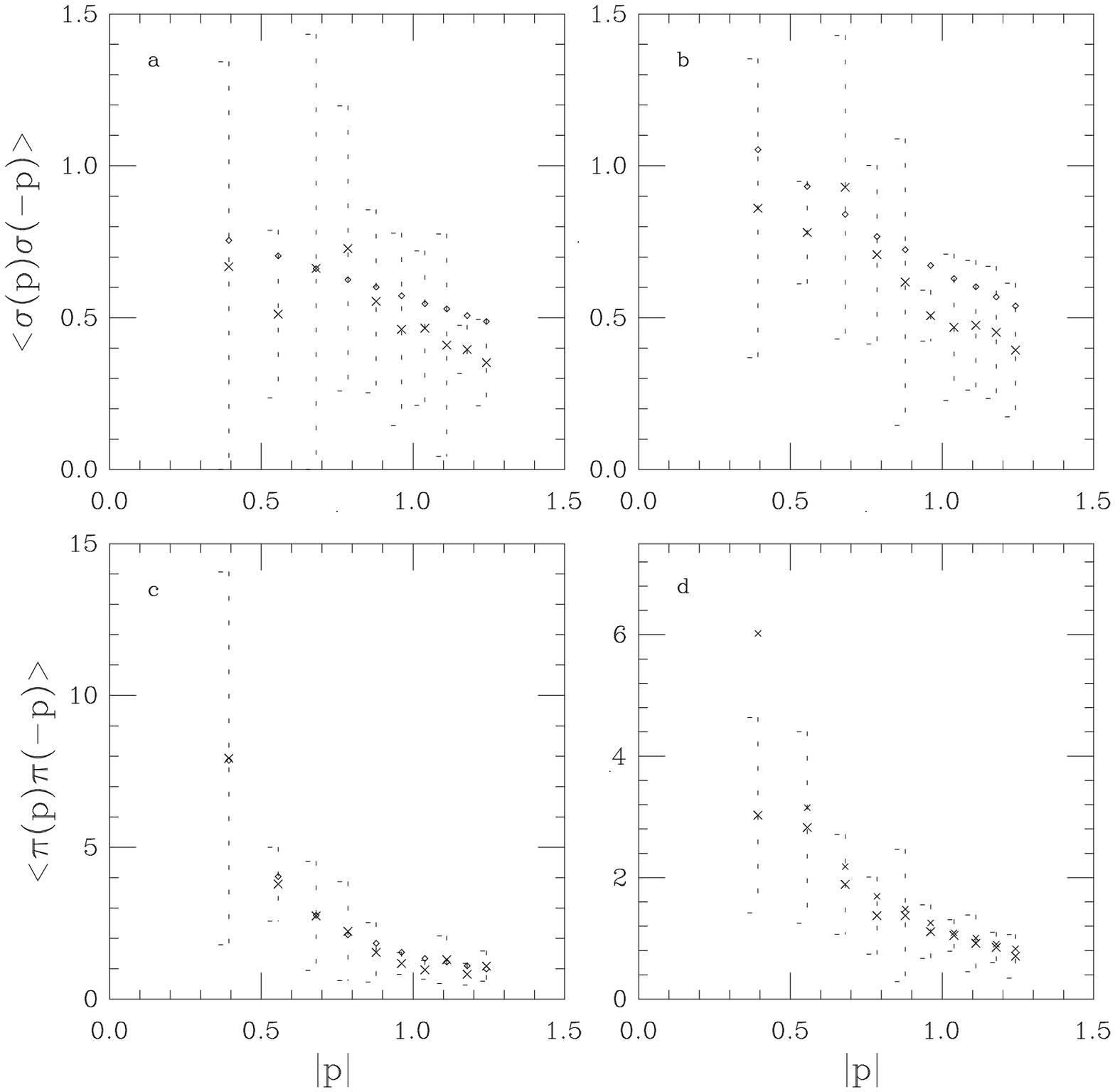}{\four}{\figsizeA}

\medskip
\noindent {\bf VI}
\smallskip

The numerical and the leading order large N results are in good
agreement, indicating that the $1/N$ corrections are small for the
quantities we were able to measure.  The first indication that the
model at $N=2$ has same type of behavior as at $N=\infty$ appears in
Figure 3. There we plot $m_q$ and $m^{\prime}_q=\{-N <\bar\Psi\Psi> /
(2 \beta_1)\} + m_0$ vs. $\beta_1$ as determined from the numerical
simulation.  At large $N$ the gap equation gives $m_q = m^{\prime}_q$.
This relation is satisfied nicely. Furthermore, the agreement with
large $N$ of dynamically determined quantities vs. bare quantities is
also quite good and helps us to get oriented in the bare parameter
space. Figure 3 is an example of this. Some more results that
demonstrate this agreement are presented in \cite{Bitar-Vranas}.

\figure{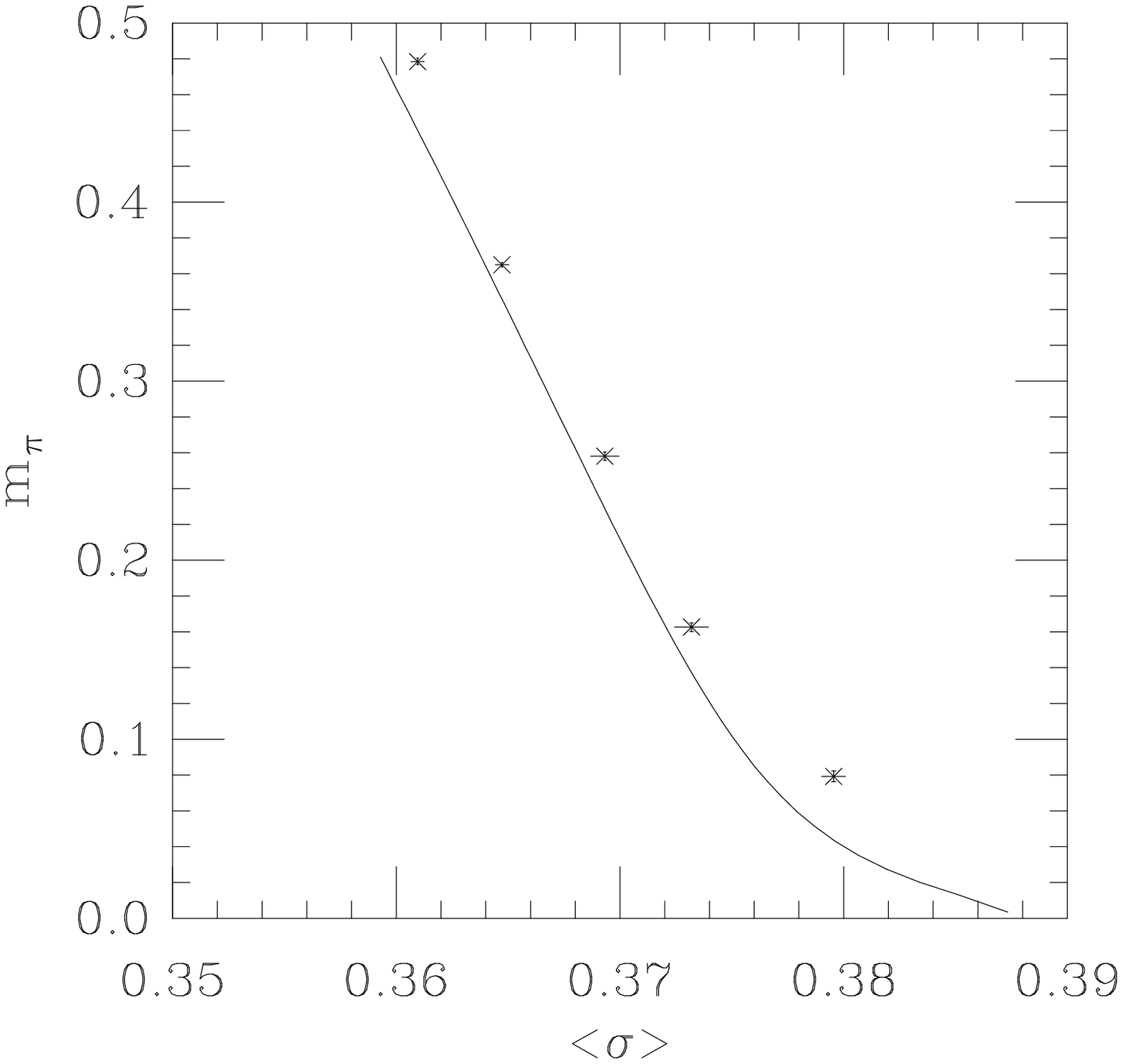}{\five}{\figsizeA}

The important comparison with large $N$ that will help us get a feel
for the size of the $1/N$ corrections comes from comparisons of
dynamically determined quantities vs. other dynamically determined
quantities. In particular we exchange one of the bare parameters for
$\langle \sigma \rangle$.  In Figure 4 we present the $\sigma$ and
$\bfm{\pi}$ propagators in momentum space for ten small momenta and
$r=0$. It is from these figures that we would have to extract the pion
wave function renormalization constant $Z_\pi$. As it can be seen the
large $N$ prediction for the same $\langle \sigma \rangle$ as the one
measured in the simulation is in good agreement with the numerical
results. The large $N$ predictions in Figures 4a, 4b, 4c, and 4d
``fit" the numerical results with $\chi^2$ per degree of freedom
$0.38$, $0.67$, $0.32$, $0.42$ respectively. This means that the pion
decay constant in lattice units $f_\pi= \langle \sigma \rangle \sqrt{N
\over Z_\pi}$ \cite{Ebert-Volkov} as a function of $\langle \sigma
\rangle$ has very small $1 / N$ corrections.  Another dynamically
determined quantity that agrees very well with the large $N$
prediction when plotted vs. $\langle \sigma \rangle$ is the pion mass
$m_\pi$. This plot is shown in Figure 5.  This figure suggests that
the $1/N$ corrections to $m_\pi$ are very small.

At large $N$ we find that on a finite lattice $m_q$ can become zero
only if $\beta_1$ becomes infinite (free field).  On the other hand we
find that on a finite lattice $m_\pi$ can be made to vanish by
adjusting the bare parameter $m_0$. Indeed we can see from Figure 5
that $m_\pi$ can be made very small.  In fact it turns out that the
value of $m_0$ where this happens is predicted by large $N$ quite
well.

As it has already been discussed in {\bf V}, for $r\neq0$ the $\sigma$
mass is of order cutoff and therefore very heavy to be able to measure
from the decay of the $\sigma -\sigma$ correlation function. However
for $r=0$ one would expect to be able to measure $m_\sigma$. As we
will discuss in {\bf VII} below this is not possible with the lattice
sizes accessible to us. This is unfortunate since $m_\sigma$ is
another very important quantity. However, we expect the size of the
$1/N$ corrections of $m_\sigma$ to be similar to the ones of $m_\pi$
and therefore very small. Also, we did not measure the $\sigma$ width
in the numerical simulation, but, as discussed in {\bf I}, we expect
the $1/N$ corrections to be large.

\medskip
\noindent {\bf VII}
\smallskip

On finite lattices the zero momentum modes of the quarks affect the
inversion speed of the CG algorithm and introduce finite size effects.
To leading order at large $N$ the matrix $M^\dagger M$ of eq.
\ref{PartFun2} is diagonal in momentum, spin, flavor, and color
spaces.  The smallest eigenvalue of this matrix is $m_q^2$ and
corresponds to the $p=0$ matrix element. This in turn corresponds to
the zero momentum mode of the quarks which from now on we will simply
call ``zero modes''.

On a finite lattice the role of the zero modes is crucial in the
inversion of $M^\dagger M$.  For small $m_q$, the condition number of
the matrix is $4/m_q^2$ for $r=0$ and $64 r / m_q^2$ for $r=1$ and it
is clear that it depends strongly on the presence of the zero modes. A
large condition number will make the inversion of $M^\dagger M$ very
slow. Furthermore the spacing of the smallest eigenvalues behaves like
$1 / L^2$ and for larger lattices the inversion times will rapidly get
worse. An important observation can be made by noticing the
dependence of the condition number on $r$. This suggests that
performing the simulation with smaller $r$ will yield a quite faster
inversion. It is possible that this may also be the case for QCD.

But the unwelcomed effect of the zero modes on a finite lattice is not
limited to large inversion times. Because on a finite lattice their
effects are not suppressed by the measure but instead by an inverse
volume factor, it turns out that in certain cases they severely obscure
the physics. By simply looking at the numerical results in Figure 3 we
would not only be unable to estimate the critical point but also we
would be unable to see any indication of a phase transition. The large
$N$ result on the same size lattice also has the same problems. As we
increase the lattice size in the large $N$ calculation (solid lines
from top to bottom) we see that a picture of an order parameter slowly
materializes. At $ L_x=64 $, $ L_t=64 $ a fairly good prediction of
the large $N$ infinite volume critical point is achieved. If we now do
the same large $N$ calculation but neglect from the momentum sums the
zero modes, we obtain as a result the two dotted lines for $L_x=8$,
$L_t=16$ and $L_x=16$, $L_t=16$ (from left to right). We see that
neglecting the zero modes on an $L_x=8$, $L_t=16$ lattice gives very
similar results as the ones obtained on a $L_x=64$, $L_t=64$ lattice
with the zero modes included.

As mentioned earlier although one would expect to be able to measure
the $\sigma$ mass in the $r=0$ case we were not able to do so. Large
$N$ provides an explanation of this unexpected problem.  If we plot
the real part of the inverse $\sigma$ propagator for a finite
lattice and set the external momentum to $q=\{i
m_\sigma,0,0,0\}$ we can obtain the sigma mass at the zero of
this function. We find that because of the presence of a discontinuity
we do not obtain a root at all. A root will eventually be obtained
but not until $m_\sigma$ becomes very heavy \cite{Bitar-Vranas}. The
presence of this discontinuity is again due to the zero modes.

There are some very interesting issues relevant to lattice work that
have not been considered in this paper. It would be important to
calculate the three and four point vertices and therefore be able to
calculate scattering amplitudes and their departure from low energy
behavior as well as the $1/N$ corrections to the width.  It would also
be interesting to study the NJL model at finite temperature and
investigate the finite temperature transition in connection with the
approach to the chiral limit. Finally it would be important to include
vector meson couplings (see, for example, \cite{Ebert-Volkov},
\cite{Ebert-Reinhard}) and confirm that the vector meson masses scale
appropriately and do not become of the order cutoff as the $\sigma$
particle does.

\section*{Acknowledgments}

We would like to thank U.M. Heller and R. Edwards for useful
discussions concerning the subject of this paper. This research was
supported in part by the DOE under grant $\#$ DE-FG05-85ER250000 and
$\#$ DE-FG05-92ER40742.


\if \epsfpreprint N

\eject

\section* {Figure Captions.}

\noindent{\bf Figure 1:} \one

\noindent{\bf Figure 2:} \two

\noindent{\bf Figure 3:} \three

\noindent{\bf Figure 4:} \four

\noindent{\bf Figure 5:} \five

\fi

\end{document}